# Disentangling complex current pathways in a metallic Ru/Co bilayer nanostructure using THz spectroscopy

Nicolas S. Beermann, Savio Fabretti, Karsten Rott, Hassan A. Hafez, Günter Reiss, and Dmitry Turchinovich[*]

*Fakultät für Physik, Universität Bielefeld, Universitätsstr. 25, 33615 Bielefeld, Germany*

[*]Corresponding author: dmtu@physik.uni-bielefeld.de

Many modern spintronic technologies, such as spin valves, spin Hall applications, and spintronic THz emitters, are based on electrons crossing buried internal interfaces within metallic nanostructures. However, the complex current pathways within such nanostructures are difficult to disentangle using conventional experimental methods. Here, we measure the conductivity of a technologically relevant Ru/Co bilayer nanostructure in a contact-free fashion using THz time-domain spectroscopy. By applying an effective resistor network to the data, we resolve the complex current pathways within the nanostructure and determine the degree of electronic transparency of the internal interface between the Ru and Co nanolayers.

Spintronics is a rapidly developing field of modern science and technology.[1,2] The majority of spintronics applications, such as spin valves, spin transistors or spintronic THz emitters are based on functional multi-layer structures comprising different metallic nanofilms.[3,4] The electric currents within such structures assume complex pathways. However the role of internal metal-metal interfaces in total current distribution within the overall structure remains a subject of intensive research.[5–7] The straightforward characterization of complex multi-layer structures, based on applying electrical contacts, is not always reliable as the contact effects may influence the measurement itself. Here the typical problems arise e.g. from the contact resistance effects in characterization of field-effect transistor structures[8], or from the spin accumulation at the contacts complicating the measurements on spin valves.[9,10] Therefore, sensitive non-contact methods are required for careful studies of electrical transport in complex nanostructures.

THz time-domain spectroscopy (THz-TDS) is a contact-free, all-optical spectroscopy method, which is ideally suited for studies of electrical transport on a nanoscale (see e.g.[11–13]). Here, we applied THz-TDS to studies of conduction in technologically-relevant metallic bilayer Ru/Co nanostructures of different thicknesses [14–16], and of their individual constituent metals Ru and Co. Such structures find applications, among others, in spin-torque devices[17,18], as well as in structures featuring magnetic skyrmions.[19–21] All the THz conductivity spectra measured in our work could be well described by the Drude model of free carrier conduction, and the conductivity of bilayers was analyzed based on an effective resistor network model. As a result, we were able to access the complex electrical current distribution within the Ru/Co bilayers, and establish the coefficient of electronic transparency of an internal Ru/Co interface.

Our metallic structures were produced by DC-magnetron sputtering on 500 μm thick (100) MgO substrates. The base pressure during this process was $3 \times 10^{-7}$ mbar while the metallic layers were deposited at an argon pressure of 3.5×10⁻³ mbar. The sputtering rate of Co was 1.4 Å/s and for Ru it was 0.8 Å/s. Our sample set consisted of a clean MgO reference sample; a 10 nm thick Ru layer; a series of 5, 10, 20, and 40 nm - thick Co layers; and a series of Ru/Co bilayers. In our Ru/Co bilayer samples, the Ru layer thickness was kept constant at 10 nm, and the Co layer thickness varied as 5, 10, 20, and 40 nm. All our samples were additionally capped with a 2 nm MgO top layer, in order to protect the metallic layers from oxidation.

We measured the THz conductivity of our samples in a standard normal-incidence transmission-mode configuration in the frequency range $f$ = 0.3 – 2.5 THz, using a commercial THz-TDS spectrometer (Teraflash from Toptica Photonics AG). Normal incidence of the THz beam onto the sample results in net in-plane THz-driven electric current within the metallic nanofilm, enabled by its conductivity, and obeying Ohm's law. All our experiments were performed at room temperature, in a dry nitrogen atmosphere.

First we have characterized the (100) MgO substrate. We found no measurable THz absorption, while the refractive index showed weak frequency dependency. The refractive index dispersion in our spectroscopy window of $f$ = 0.3 – 2.5 THz could be well fitted by a function $n_\text{s}(f) = 3.21 - 0.066\,f + 0.027\,f^2$, where $f$ is frequency in THz. The THz frequency - averaged value of refractive index of (100) MgO was $n_\text{s} = 3.19 \pm 0.04$. Further, the complex-valued THz field transmission spectrum $\tilde{T}(f)$ of the metallic film samples referenced to the bare substrate was measured. Our samples represent conductive films with the thickness much smaller than the wavelength of the incident THz light, deposited on a non-conductive substrate. THz transmission through such a subwavelength conductive film results in multiple constructive interference of incident THz light within the film, and hence also in a uniform distribution of THz field strength $E$ along the entire thickness of the film sample. Under such an assumption, the THz transmission spectrum $\tilde{T}(f)$ through the sample is related to the complex sheet conductivity $\tilde{\sigma}_\text{s}(f)$ of the conductive film via the Tinkham equation [22–24]

$$\tilde{T}(f) = \frac{\tilde{E}_\text{sam}(f)}{\tilde{E}_\text{ref}(f)} = \frac{1+n_\text{s}(f)}{1+n_\text{s}(f) + Z_0\,\tilde{\sigma}_\text{s}(f)}. \qquad (1)$$

Here $\tilde{E}_\text{ref}(f)$ and $\tilde{E}_\text{sam}(f)$ are the frequency-dependent complex-valued electric field amplitudes of THz pulses transmitted through the bare substrate and through the metallic film on the substrate, respectively; $n_\text{s}(f)$ denotes the previously determined refractive index of the substrate, and $Z_0 = 377\,\Omega$ is the impedance of free-space. The complex conductivity $\tilde{\sigma}$ is related to the sheet conductivity via $\tilde{\sigma}_\text{s}(f) = \tilde{\sigma}(f)\,d$, where $d$ is the thickness of the conductive film.

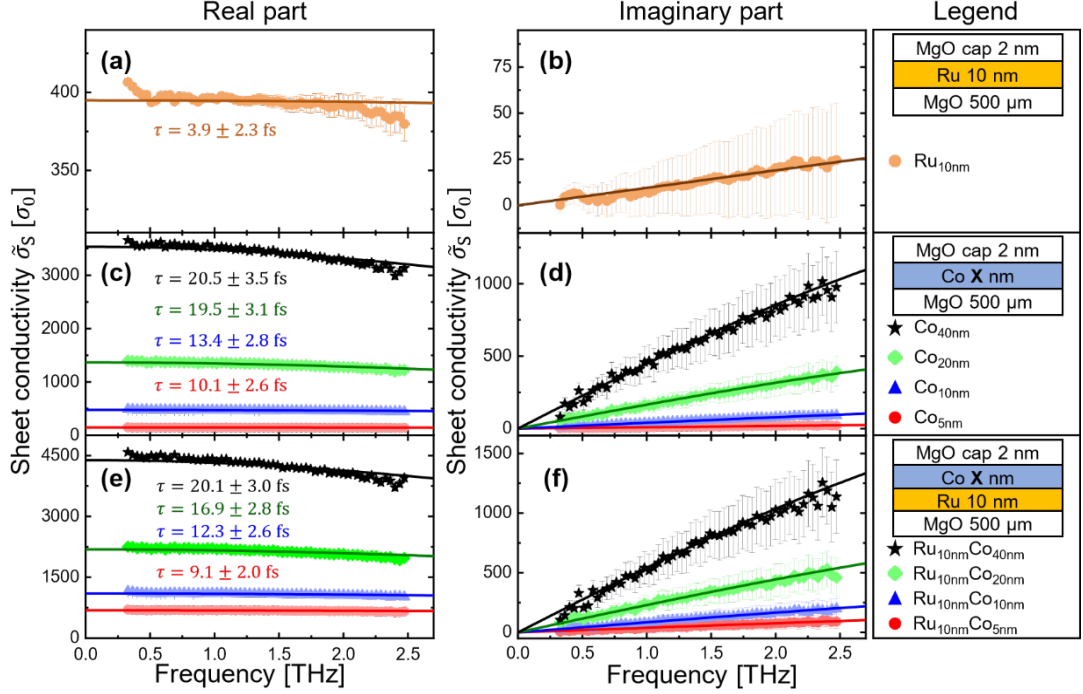

**Fig.1**: *Symbols: THz conductivity spectra for Ru, Co and Ru/Co films, normalized to the quantum conductance unit $\sigma_0 = \frac{2e^2}{h} \approx 7.75 \times 10^{-5}$ S and split into real ((a), (c) and (e)) and imaginary parts ((b), (d) and (f)). The solid lines represent the Drude fits. Established electron momentum scattering times are indicated in the Figure.*

Fig.1 shows the measured complex-valued sheet conductivity spectra of all our metallic samples: Ru and Co nanofilms, as well as Ru/Co bilayer structures of different overall thicknesses. We note that due to a minor difference in the thicknesses of reference and sample substrates, which can be established by measuring the multiple-reflection etalon signals in THz-TDS (see e.g.[25] and Supplementary information), a frequency-dependent correction to the phase of the transmission spectrum $\tilde{T}(f)$ was introduced in our data analysis. The uncertainty in this phase correction translates into the frequency-increasing error bar of the imaginary part of the measured conductivity, which in turn is strongly dependent on the phase of $\tilde{T}(f)$.

All our measured conductivity spectra could be reasonably well fitted by the classical free-carrier conduction Drude model (Fig.1 solid lines).

$$\tilde{\sigma}_s(\omega) = \tilde{\sigma}(\omega)\, d = \frac{\sigma_{DC}}{1 - i\,\omega\,\tau}\, d, \qquad (2)$$

where $\omega = 2\pi f$ is the angular frequency. The DC-conductivity $\sigma_{DC}$ and the electron momentum scattering time $\tau$ for all our samples were obtained from the Drude fits, and are summarized in Fig. 2. The electron momentum scattering time $\tau$ is strongly dependent on the metal type and on the sample thickness, and ranges from ~4 fs to ~20 fs. We find saturating growth of both the DC-conductivity $\sigma_{DC}$ and the scattering time $\tau$ with increasing sample thickness.

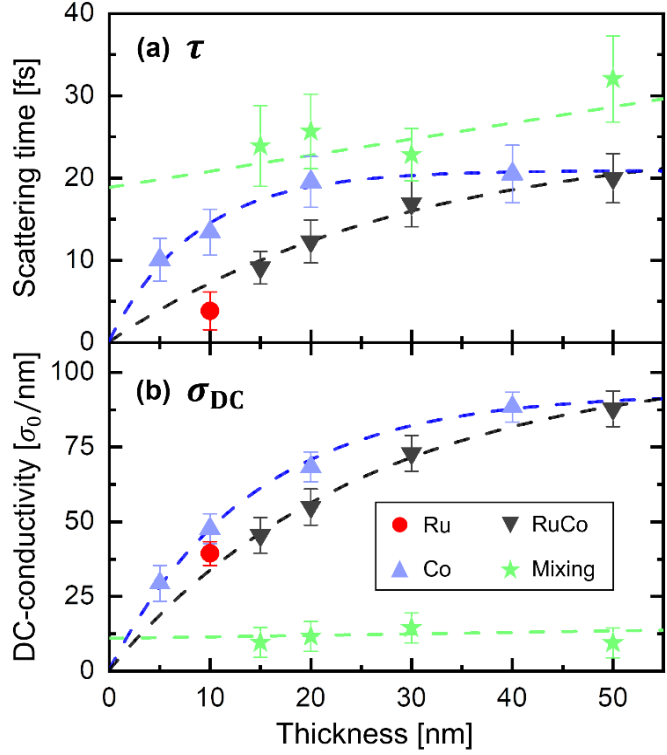

**Fig.2**: *(a) Electron momentum scattering time τ and (b) DC-conductivity $\sigma_{DC}$, established from the Drude fits parameters for all metallic thin films, as well as for the mixing contribution, as function of the total sample thickness. $\sigma_0$ corresponds to the quantum conductance unit of $7.75 \times 10^{-5}$ S. The dashed lines are guides to the eye.*

Our data from Fig. 2 correlate rather well with available literature. For Co, the electron momentum scattering time $\tau$ for film thicknesses of $d \geq 20$ nm tends to reach the steady value of $\tau \approx 20$ fs, which agrees very well with theoretical results from [26], predicting the value of $\tau$ = 21.2 fs for bulk Co. This saturating dependency of the electron momentum scattering time $\tau$ on the film thickness manifests the transition to bulk conduction regime in Co for sample thickness exceeding 20 nm. This thickness should be of the order of the electron mean free path in bulk Co, where the electron transport is no longer affected by scattering on external interfaces. Similarly, the DC-conductivity $\sigma_{DC}$ of Co films and Ru/Co bilayers depends on the sample thickness, and shows a similar saturation with increasing layer thicknesses. This is in good agreement with conventional transport measurements on Co thin films. [27]

One can now investigate the current distribution within the bilayer Ru/Co structures based on the measured individual conductivities of the constituent metals Ru and Co (Fig. 1 (a-d)), and of the Ru/Co bilayers (Fig. 1 (e-f)). Under the above-mentioned reasonable assumption of uniform THz electric field distribution along the entire thickness of the bilayer, we first tested the hypothesis of two parallel, non-mixing THz-driven currents flowing independently in Ru and Co sections of a Ru/Co bilayer. Such non-mixing conduction in a bilayer would imply an internal metal-metal Ru/Co interface being electronically non-transparent. This situation is described by an effective circuit of two parallel-connected resistors $R_{Ru}$ and $R_{Co}$, representing the Ru and Co films respectively. In this case, the sheet conductivity of the entire Ru/Co bilayer structure should be expressed as $\tilde{\sigma}_{s,Ru/Co} = \frac{1}{R_{Ru/Co}} = \frac{1}{R_{Ru}} + \frac{1}{R_{Co}} = \tilde{\sigma}_{Ru}d_{Ru} + \tilde{\sigma}_{Co}d_{Co}$, where $\tilde{\sigma}_{Ru}$ and $\tilde{\sigma}_{Co}$ are the conductivities of Ru and Co layers, and $d_{Ru}$ and $d_{Co}$ are their thicknesses, respectively. When comparing thus calculated values of $\tilde{\sigma}_{s,\text{Ru/Co}}(f)$ with the measured ones, shown in Fig.1 (e-f), we have found that the measured values exceeded the calculated ones in all cases, i.e. for all thicknesses of Ru/Co bilayers. Consequently, we conclude that the hypothesis of a non-transparent internal interface between Ru and Co nanolayers is false. Since the combined conductance of a Ru/Co bilayer turns out to always exceed that of the sum of individual conductances of its constituent Ru and Co nanofilms, we conjecture that an additional in-plane current channel should exist in the bilayer, namely a mixing current flowing in both Ru and Co parts of the structure. The existence of such a current will then imply a (partial) electronic transparency of the internal Ru/Co interface, which parameter can now be possibly determined.

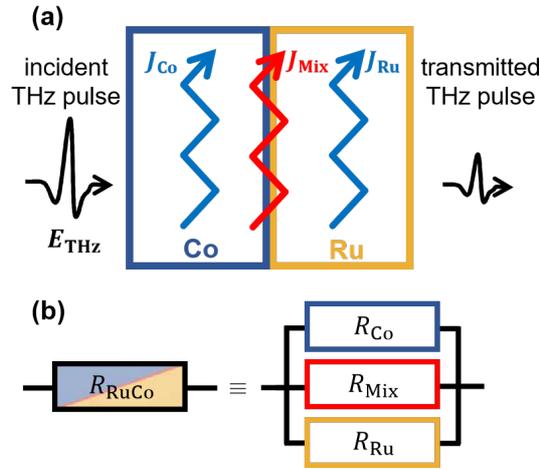

**Fig.3**: *(a) Schematic picture of the Ru/Co metal bilayer, indicating non-mixing $J_{Co}$ and $J_{Ru}$ and mixing $J_{Mix}$ currents, driven by incident THz fields. (b) The corresponding effective resistor network describing the current pathways above.*

We now describe the in-plane conduction in our metallic bilayer as illustrated in Fig. 3(a). The incident THz field drives three parallel in-plane currents in the bilayer structure: two non-mixing currents $J_{\text{Co}}$ and $J_{\text{Ru}}$ and a third, mixing current $J_{\text{Mix}}$ representing the electronic transparency of the internal interface between the Co and Ru nanofilms. Consequently, the conduction in the bilayer can be described as a parallel connection of not two, but three corresponding resistors, as shown in Fig. 3(b). Accordingly, the sheet conductivity of a bilayer is now expressed as

$$\tilde{\sigma}_{s,\text{RuCo}} = \tilde{\sigma}_{s,\text{Ru}} + \tilde{\sigma}_{s,\text{Co}} + \tilde{\sigma}_{s,\text{Mix}}, \qquad (3)$$

where the first three terms $\tilde{\sigma}_{s,\text{RuCo}} = \tilde{\sigma}_{\text{RuCo}} d_{\text{RuCo}}$, $\tilde{\sigma}_{s,\text{Ru}} = \tilde{\sigma}_{\text{Ru}} d_{\text{Ru}}$ and $\tilde{\sigma}_{s,\text{Co}} = \tilde{\sigma}_{\text{Co}} d_{\text{Co}}$ are determined experimentally. Hence, the mixing conductivity term $\tilde{\sigma}_{s,\text{Mix}}$, dependent on the electronic transparency of an internal Ru/Co interface, can be now isolated, and is shown in Fig. 4 as function of THz frequency and bilayer thickness. We observe the overall increase in this mixing sheet conductivity $\tilde{\sigma}_{s,\text{Mix}}$ for the Ru/Co bilayer with increase in the sample thickness, exhibiting certain saturation for the real parts of $\tilde{\sigma}_{s,\text{Mix}}$ for bilayer samples with the thickest Co parts of 20 nm and 40 nm.

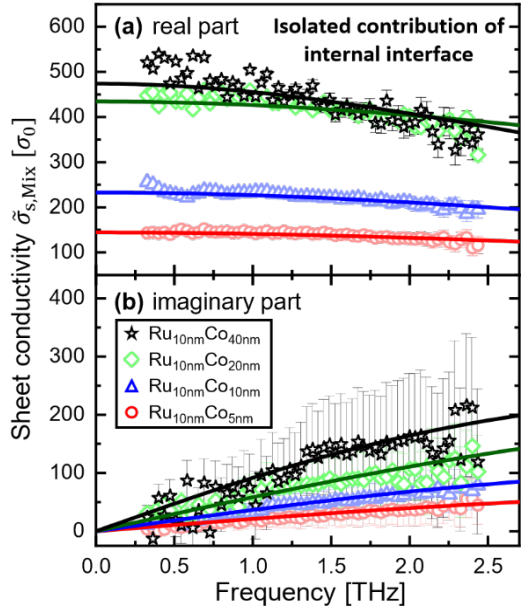

**Fig.4**: *Complex conductivity spectra for the mixing contribution $\tilde{\sigma}_{s,Mix}$ of the Ru/Co bilayer, for **(a)** real and **(b)** imaginary parts. The solid lines represent Drude fits.*

This sheet conductivity $\tilde{\sigma}_{s,Mix}$ is again well described by the Drude model, with the fit parameters represented in Fig. 2 and denoted as "Mixing". Interestingly, the DC conductivity of the mixing contribution, describing the magnitude of the mixing current within the bilayer structure, is found to be essentially independent of the sample thickness, with a mean value of $\sigma_{DC,Mix} = 11.3 \pm 4.1\ \sigma_0/\text{nm}$. We note, that this value is considerably lower than that of individual constituent metals Ru and Co, for all sample thicknesses measured. The electron momentum scattering time $\tau$, related to the mixing conductivity term, lies in the range $\tau \approx 20$ - 30 fs for all the thicknesses of Ru/Co structure. This scattering time, representing the average duration of an elementary mixing current, is longer than that of the individual constituent metals Ru and Co, and of the Ru/Co bilayer of any thickness. This lower DC conductivity combined with longer momentum scattering time for the electrons assigned to the mixing contribution in overall conductivity of the bilayer, may indicate the following. The presence of internal metal-metal Ru/Co interface may provide for lower electron scattering rate as compared to the bulk of the individual constituent metals. At the same time, the effective Fermi surface for such electrons may also be reduced, leading to a lower density of conduction electrons participating in this mixed current.

Now, based on the established sheet conductivities, corresponding to the currents $J_{Co}$, $J_{Ru}$, and $J_{Mix}$ (see Fig. 3 and Eq. 3), we define the coefficient $\tilde{t}_{Int}$, describing the electronic transparency of the internal Ru/Co interface, and hence the distribution of mixing and non-mixing currents within the bilayer nanostructure. This coefficient $\tilde{t}_{Int}$ describes the relative strength of the mixing current in relation to the sum of non-mixing currents flowing within the bilayer:

$$\tilde{t}_{Int} = \frac{J_{Mix}}{J_{Ru}+J_{Co}} = \frac{\tilde{\sigma}_{s,Mix}}{\tilde{\sigma}_{s,Ru}+\tilde{\sigma}_{s,Co}}. \quad (4)$$

Here, Ohm's law $J = \tilde{\sigma}_s E$ connects the currents $J$ and the corresponding sheet conductivities $\tilde{\sigma}_s$, with THz electric field strength $E$ being uniform within the whole metallic nanostructure as explained above. The established interface transparency coefficient $\tilde{t}_{Int}$ is shown in Fig. 5. We find, that $\tilde{t}_{Int}$ is predominantly real-valued, and is almost frequency-independent. The frequency-averaged value of $\text{Re}\{\tilde{t}_{Int}\}$ shows dependency on the thickness of the Ru/Co bilayer: while for the structures with an overall thickness of 15 – 30 nm this value remains almost constant at about $\text{Re}\{\tilde{t}_{Int}\} \approx 0.25$, for the thickest sample of 50 nm it drops dramatically to $\text{Re}\{\tilde{t}_{Int}\} \approx 0.12$, i.e. by over a factor of two.

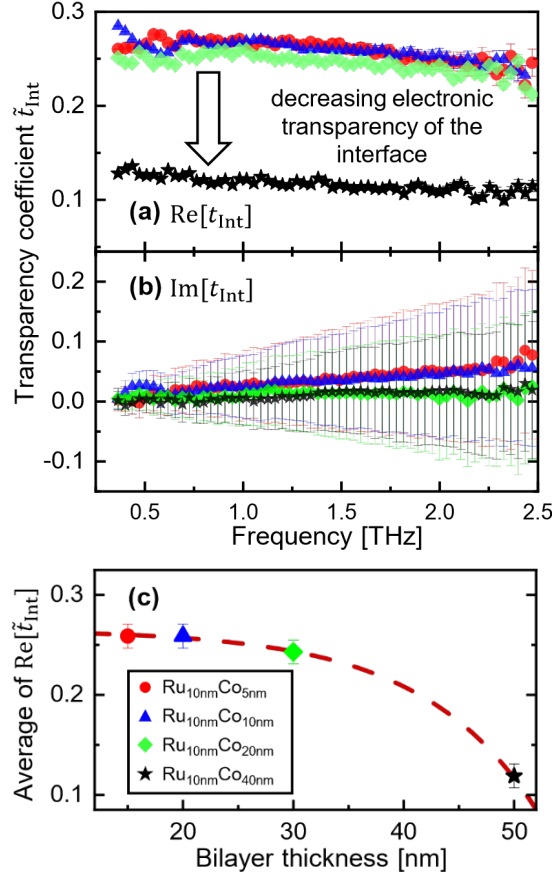

**Fig.5**: *Coefficient of electronic transparency of internal Ru/Co interface $\tilde{t}_{Int}$. (a) Frequency-dependent real and imaginary part and (b) frequency-averaged real part, for different sample thicknesses. Dashed line – guide to the eye.*

Since the interface transparency coefficient $\tilde{t}_{Int}$ quantifies the relative strength of the mixing current within the nanostructure, and hence the prominence of an internal interface in the total electrical transport, it is clear that with the growth of the overall structure thickness the role of an internal interface should also diminish. Its value should therefore decrease with further increase in the overall thickness of a bilayer, and will reach zero at a thickness, when the conduction in the bilayer can be well described by only two non-mixing bulk currents independently flowing in each part of the structure.

In summary, we studied the in-plane conductivity of Ru/Co bilayer nanostructures and of thin films of individual constituent metals Ru and Co using THz-TDS. The conductivities of all measured samples are well described by the classical Drude model. Using the model based on an effective resistor network, we managed to disentangle the complex current pathways within the Ru/Co bilayer nanostructure, and establish the coefficient of electronic transparency of the internal metal-metal Ru/Co interface, which was found to have a strong dependency on the bilayer thickness. Our method introduced here on the example of Ru/Co nanostructures, can serve the basis for a general protocol of quantitative analysis of buried internal interfaces in many different material structures, not only metallic.

Supplementary Material: See the supplementary information for the details on frequency-dependent phase correction to the transmission spectra.

We acknowledge the financial support from European Union's Horizon 2020 research and innovation programme (grant agreement N° 964735 EXTREME-IR), and Deutsche Forschungsgemeinschaft (DFG) within the project 468501411–SPP2314 INTEGRATECH under the framework of the priority programme SPP2314 – INTEREST